# Ballistic transport spectroscopy of spin-orbit-coupled bands in monolayer graphene on WSe$_2$

Qing Rao,[1]† Wun-Hao Kang,[2]† Hongxia Xue,[1] Ziqing Ye,[3] Xuemeng Feng,[3] Kenji Watanabe,[4] Takashi Taniguchi,[4] Ning Wang,[3] Ming-Hao Liu,[2]* and Dong-Keun Ki[1]*

[1]Department of Physics, The University of Hong Kong, Pokfulam Road, Hong Kong, China
[2]Department of Physics, National Cheng Kung University, Tainan 70101, Taiwan
[3]Department of Physics and Center for Quantum Materials, The Hong Kong University of Science and Technology, Clear Water Bay, Kowloon 999077, Hong Kong, China
[4]National Institute for Materials Science, Namiki 1-1, Tsukuba, 305-0044, Ibaraki, Japan

†Equal contributions.
*Corresponding author. Email: minghao.liu@phys.ncku.edu.tw, dkki@hku.hk

**Van der Waals interactions with transition metal dichalcogenides was shown to induce strong spin-orbit coupling (SOC) in graphene, offering great promises to combine large experimental flexibility of graphene with unique tuning capabilities of the SOC that can rotate spin by moving electrons or vice versa. Here, we probe SOC-driven band splitting and electron dynamics in graphene on WSe$_2$ by measuring ballistic transverse magnetic focusing. We found a clear splitting in the first focusing peak whose evolution in charge density and magnetic field is well reproduced by calculations using SOC strength of ~13 meV and no splitting in the second peak that indicates stronger Rashba SOC. A possible suppression of electron-electron scatterings was also found in temperature dependence measurement. Further, we found that Shubnikov-de Haas oscillations exhibit SOC strength of ~3.4 meV, suggesting that it probes different electron dynamics, calling for new theory. Our study demonstrates an interesting possibility to exploit ballistic electron motion pronounced in graphene for emerging spin-orbitronics.**

The interfacial interactions with semiconducting transition metal dichalcogenides (TMDCs) have shown to be highly efficient to induce strong spin-orbit coupling (SOC) in graphene(*1-3*). For monolayer graphene, it was theoretically predicted to have two distinctive terms, one that couples out-of-plane spin and valley degrees of freedom (referred to as a spin-valley Zeeman term $\tau_z s_z$) and another that couples in-plane spin and sublattice degrees of freedom, similar to the Rashba term ($\tau_z \sigma_x s_y - \sigma_y s_x$), as follows.

$$H = H_0 + \Delta \sigma_z + \lambda \tau_z s_z + \lambda_R (\tau_z \sigma_x s_y - \sigma_y s_x), \quad (1)$$

where $H_0$ is the graphene's Dirac Hamiltonian, $\boldsymbol{\sigma} = (\sigma_x, \sigma_y, \sigma_z)$ is a Pauli matrix vector that acts on the sublattice degree of freedom in graphene, $\boldsymbol{s} = (s_x, s_y, s_z)$ is a Pauli matrix vector that acts on spin, and $\tau_z = \pm 1$ identifies two different valleys in graphene ($\Delta \approx 0$ due to a large lattice mismatch between the graphene and TMDCs)(*2, 4-7*). Both SOC terms induce band splitting in graphene and their strengths ($\lambda$ and $\lambda_R$) can be further tuned by an electric field perpendicular to the layers(*8-13*), twisting (*14, 15*), or by pressure(*16*).

Combined with high electron mobility and large experimental flexibility of graphene, such a strong interface-induced SOC makes the graphene on TMDCs ideal for ballistic spin-orbitronics where ballistic electron motion is used to control or detect electron spin(*17-22*). It is particularly interesting as graphene has shown pronounced ballistic transport effects with large tunability, such as transverse magnetic focusing (TMF)(*23-25*), Veselago lensing(*26, 27*), Fabry-Pérot interference(*28, 29*), and ballistic snake states(*30, 31*) among many others. They have also shown unique features originating from the relativistic nature of Dirac electrons(*23-31*). However, a vast number of the previous studies on graphene-TMDC heterostructures have focused on detecting spin relaxation due to electron scattering rather than the ballistic motion(*1-3, 7, 16, 32-37*). Moreover, only few studies have found a direct evidence for the SOC-induced band splitting by measuring beatings in Shubnikov-de Hass (SdH) oscillations (for both mono- and bilayer graphene)(*3, 10, 38*) or tracing changes in quantum capacitance (for bilayer graphene only)(*11*). Not only to understand the effect of the SOC on the electronic properties of the system such as the band topology(*39, 40*) but also to exploit its full potential on ballistic spin-orbitronics(*17-22*), it is therefore essential to demonstrate the ballistic transport in graphene on TMDCs while simultaneously probing their band structures and electron dynamics.

To fill this missing link, we employ TMF technique in monolayer graphene on WSe$_2$ as it can not only probe the SOC-induced band splitting but also investigate electron dynamics simultaneously (see Figs. 1A,B). TMF occurs when ballistic carriers injected from a narrow aperture ("injector") at the edge of the sample are subject to a small perpendicular magnetic field ($\mathbf{B} = B\hat{z}$)(*41*). Owing to the Lorentz force, the carriers follow skipping cyclotron orbits and focus on another narrow aperture ("collector") at a distance ($L$) that equals an integer multiple of $2r_c$ with a cyclotron radius $r_c = \hbar k_F/eB$, where $\hbar$ is the reduced Planck constant, $k_F$ is the Fermi momentum, and $e$ is the elementary charge. Upon sweeping magnetic fields, the collector voltage will exhibit a set of resonance peaks at certain $B$-values determined by $k_F$,

$$B_j = \pm \frac{2j\hbar}{|e|L} k_F, \quad (2)$$

where $j$ is an integer and $\pm$ represents electron and hole for the configuration shown in Fig. 1A). This enables the detection of Fermi surface configurations(*41-43*). In the systems with multiple bands(*23*), for instance, there will be multiple sets of resonance peaks at different $B$-values from which one can deduce their band structures. Moreover, TMF can also be used to study or control electron dynamics as charge carriers follow skipping cyclotron orbits during the process. In 2D electron gas (2DEG) systems with SOC, the TMF was indeed used to probe spin-orbit split bands and extract SOC strength by studying the separation of the peaks(*17, 19, 21*), deduce spin polarization by comparing their heights(*17*), or focus spin-polarized current by controlling ballistic electron motion(*17, 19, 20*). Fig. 1B shows the energy band structure of the monolayer graphene on WSe$_2$ calculated using the Hamiltonian Eq. (1) that exhibits two spin-orbit split bands, $S_+$ and $S_-$, and the simulated two-point conductance spectra exhibiting multiple focusing peaks (see **Supplementary Materials** for simulation details)(*44, 45*).

**Sample characterization**

To study the ballistic TMF effect, we used a dry pick-up and transfer technique to assemble a stack of hexagonal boron-nitride (h-BN), monolayer graphene, and multilayer WSe$_2$ such that the graphene is protected from the harsh chemical environments in the following nanofabrication process(*46*). Standard electron beam lithography and lift-off were carried out to make Hall bar devices on 285-nm-thick SiO$_2$ substrate with doped silicon underneath used as a gate to control charge density $n$ (see Fig. 1A and **Materials and Methods** for fabrication details). Electron transport through the fabricated devices was measured in a 1.5 K variable temperature insert with 14-T superconducting magnet using a standard low-frequency AC lock-in technique (see **Materials and Methods**). Fig. 1C shows the carrier density dependences of the four-probe resistances of two different samples 1 and 2 measured at 1.5 K, exhibiting high quality with carrier mobility of 200,000 ~ 400,000 cm$^2$V$^{-1}$s$^{-1}$. Especially on the hole side, the symmetry broken quantum-Hall states were observed at magnetic fields as low as ~3 T (marked by black triangles in Fig. 1D). This indicates higher hole mobility in our sample than on the electron side. It is also consistent with the observation of the larger negative non-local Hall resistance $R_{ae,bf}$ on the hole side (the inset of Fig. 1C) originating from the ballistic electron motion(*47*) ($R_{\alpha\beta,\gamma\delta} \equiv V_{\gamma\delta}/I_{\alpha\beta}$ which refers to the resistance measured by sending a current from contact $\alpha$ to $\beta$ and measuring voltage between contacts $\gamma$ and $\delta$). Having such a high mobility—equivalently, a long mean free path—is important to resolve the small splitting of the focusing peak expected theoretically (Fig. 1B, right).

**Transverse magnetic focusing spectra**

The TMF signal ($R_{nl} = V_{cj}/I_{ai}$) is measured in a non-local configuration upon varying $n$ and $B$ as depicted in Fig. 1A. Figs. 2A,B show the resulting maps of $R_{nl}(n, B)$ from the sample 1 and 2, respectively. Both exhibit similar TMF spectra and their evolutions in $n$ and $B$. Overall, the positions of the *j*-th TMF peak in $B$ follow the Eq. (2) with $k_F = \sqrt{\pi|n|}$, as expected for monolayer graphene. However, on the hole side (where we found the higher sample quality), we can clearly identify the splitting of the first focusing peak that evolves continuously in $n$ and $B$ and no splitting in the second (here, we only focus on the peaks that appear in all density range). Fig. 2C and the right panel of Fig. 2B further magnify the features by plotting 1D cuts $R_{nl}(B)$ of the map at different $n$ which qualitatively matches the simulation result shown in Fig. 1B. Moreover, different from the TMF measurement on pristine monolayer graphene(*23*), we found a large TMF signal constantly exceeding 100 Ω around zero density (nearly two orders of magnitude larger than the values at finite densities; see the dark red bands near zero density in the color maps shown in Figs. 2A,B). All features found in the experiment (Figs. 2A,B) match well with our expectations for the graphene with SOC and provide valuable insights about microscopic electron processes in the system as discussed below.

**Analysis on the first focusing peak**

First, the first focusing peak splits due to the SOC-induced multiple bands $S_+$ and $S_-$ in our sample (Fig. 1B). Note that such a prolonged splitting in both $n$ and $B$ has not been observed in other graphene systems without SOC(*23-25*). Moreover, we were able to fit the positions of the first focusing peaks with calculations using the SOC strength of $\lambda_{SOC} \equiv \sqrt{\lambda^2 + \lambda_R^2} = 13.9$ meV and 12.1 meV for samples 1 and 2 respectively as marked by black dotted lines in Figs. 2A,B. Figs. 3A,B further emphasize the accuracy of the fitting by plotting the average of the normalized difference between the data and the calculation $\langle \delta B^2 \rangle$ as a function of $\lambda$ and $\lambda_R$ in a color scale (a darker color indicates the smaller $\langle \delta B^2 \rangle$ so the better fitting; see the caption for more details). We note that the fitting works for any values of $\lambda$ and $\lambda_R$ as long as they satisfy $\lambda_{SOC} = 13.0 \pm 4.7$ meV for both samples, indicating that the $\lambda$ and $\lambda_R$ has a similar effect on the splitting of the first TMF peak—equivalently, the band splitting—in the density range we studied. This is expected as in the density range explored, the Fermi energy is larger than both $\lambda$ and $\lambda_R$, leading to an identical splitting in momentum, $\Delta k = 2\sqrt{\lambda^2 + \lambda_R^2}/\hbar v_F$ with Fermi velocity $v_F \approx 10^6$ m/s for both $\lambda$ and $\lambda_R$. This is consistent with the previous SdH oscillations measurements that probe electron bands(*3, 38*).

In addition, we found that the amplitude of the first split peak closer to zero $B$ is always lower than that of the second (see, e.g., Fig. 2C). Interestingly, it was found and expected in 2DEG systems with Rashba SOC when adiabatic transitions between the quantized sub-bands at the injector with width $w$ polarize electron spin(*48*). Slightly modifying the condition derived for 2DEG(*48*), we get $\frac{\Delta k}{2k_F} > \frac{3}{16}\left(\frac{\lambda_F}{w}\right)^2 \rightarrow \lambda_R > \frac{3\pi}{8}\frac{\hbar v_F \lambda_F}{w^2} \approx 0.25{\sim}0.35$ meV for our sample with Fermi wavelength $\lambda_F \approx 30{\sim}40$ nm, and $w \approx 300$ nm. The SOC strengths estimated in our study are well in the range (Figs. 3A,B). On the other hand, in graphene on TMDCs, it was predicted that the characteristic spin winding of the spin-orbit-coupled bands leads to a current-induced spin polarization(*49*) which can result in the uneven heights of the two first focusing peaks. For better understanding, we will need to carry out more sophisticated experiments, such as using ferromagnetic contacts for spin-sensitive detection under oblique magnetic fields to fine-tune Zeeman energy of the system and spin orientations of the magnetic contacts simultaneously(*50*). Nevertheless, our study shows a possibility of using TMF to detect spin polarization of the ballistic carriers.

**Analysis on the second focusing peak**

The second key finding of this study is the absence of the splitting in the second focusing peak (Figs. 2A,B) which provides more information about the nature of the SOC in the system. It first can be interpreted as the scattering of charge carriers between the spin-orbit-coupled bands $S_\pm$ at the sample edge which leads to a single peak as illustrated in the bottom inset of Fig. 2A. To confirm this origin, we have further calculated electron trajectories for the second peak with or without the inter-band transition in the sample in Fig. S1 (see more discussions in the **Supplementary Materials**). As expected, without the inter-band transition, the second focusing peak also exhibits splitting. This confirms that the absence of the splitting in the second peak originates from the scattering between the bands $S_\pm$ at the sample edge.

Interestingly, from the behavior of the second peak, we can learn more about the relative strength of the spin-valley Zeeman and Rashba SOC terms, $\lambda$ and $\lambda_R$, because the inter-band scattering at the edge depends sensitively on the spin textures of the split bands $S_\pm$. As depicted in Fig. 3C, when only spin-valley Zeeman term exists (in other words, when $\theta_{SOC} = 0$; see Fig. 3A), spins in $S_+$ ($S_-$) band are aligned up (down) in z-direction. Thus, when backscattered at the edge, the electron at the state A in the band $S_+$ will jump to the state B in the same band unless there are enough magnetic impurities to flip the spin which is unlikely in high-quality graphene samples like ours. This would lead to the splitting of the second focusing peak as illustrated in Fig. 3C. On the other hand, when Rashba term dominates (*i.e.*, when $\theta_{SOC} = \pi/2$), $S_\pm$ bands have an opposite spin winding such that the electron at the state A in the band $S_+$ will jump to the state C in the opposite band, leading to the merging of the peaks as depicted in Fig. 3C. Therefore, we can estimate that in our system, Rashba term dominates. Similarly, studies on 2DEG systems with Rashba SOC have indeed shown no splitting in the second focusing peak(*19, 21, 51*).

To further confirm our analysis, we have simulated TMF spectra for different values of $\theta_{SOC}$ in Fig. 3D. As shown in the figure, the splitting in the second peak disappears rapidly as $\theta_{SOC}$ increases from zero and becomes nearly invisible as $\theta_{SOC} \gtrsim \pi/4$, consistent with our analysis above. It is, however, worth mentioning that in the simulation, we used an ideal edge, so we may have underestimated the intervalley scattering probabilities that occur in the real sample edge with atomic defects(*52*). Although this does not influence the Rashba-dominating case as the spin winding direction remains the same for the $S_\pm$ bands in different valleys, it can affect the result shown in Fig. 3D when the spin-valley Zeeman term dominates because the spin orientation for each $S_\pm$ band becomes opposite in different valleys. Thus, the intervalley scattering can lead to the backscattering between the $S_\pm$ bands at the edge when the spin-valley Zeeman term dominates, suppressing the splitting in the second peak. Although more experimental and theoretical works are required for complete understanding of this feature, we can roughly assume that our sample has disordered edges with atomic scale defects with resonance energy near the charge neutrality(*53*), leading to the intervalley scattering rate close to or less than that of the intravalley scattering in the density range explored. In this case, we would still expect to see the splitting when $\theta_{SOC} = 0$. Thus, we believe that the absence of the second peak splitting in Figs. 2A,B indicates the stronger Rashba SOC in our system.

**Large non-local resistance and temperature dependence**

We can also explain the observed large $R_{nl}$ near zero density (Figs. 2A,B) as the presence of the spin Hall effect (SHE)(*1*) in the system. From the very weak temperature dependence of the conductance minimum at zero density (Fig. 4A), we can first confirm that this is not from the gap opening at the charge neutrality which may have given a large TMF signal as found in the gapped trilayer graphene(*23*). In contrast, we found a large non-local Hall signal $R_{nl}^H = R_{ae,cg}$ near the charge neutrality that exceed the ohmic contribution by about 230 Ω (Fig. 4B; here we used Hall probes that are further apart from those used in the inset of Fig. 1C to reduce the influence from the ballistic negative resistance). This is consistent with the SHE found in a similar system previously(*1*).

We now examine temperature dependence of the TMF spectra in Figs. 4C,D to study the electron dynamics in the system. Upon increasing temperature, we found that the amplitude of the TMF spectra decreases (Fig. 4C). This suggests enhanced electron scattering at high temperatures. To identify the main scattering mechanism in our system, we extracted the total area below the first focusing peaks $A_1$ at varying temperatures from 1.5 K to 300 K and calculated the relative scattering length from $L_S/L_0 = (\ln[A_1(1.5K)/A_1(T)])^{-1}$, which is proportional to the effective scattering time ($L_0$ is the length of the semi-circular electron trajectory corresponding to the first focusing peak)(*24, 25*). Fig. 4D shows the result exhibiting a clear $T^{-1.8}$ dependence on both electron and hole side for different charge densities. This is between electron-phonon scattering ($T^{-1}$) and electron-electron (*e-e*) scattering ($T^{-2}$), indicating that although the *e-e* scattering is dominant in our sample, the WSe$_2$ may have screened some of the *e-e* interactions due to its large dielectric constant ($\varepsilon_0 \approx 7.9$) compared with h-BN ($\varepsilon_0 \approx 3.8$)(*54*). This suggests a possibility to use the WSe$_2$ to study the effect of SOC on *e-e* or electron-hole interaction phenomena, such as viscous charge transport(*55-57*), electron-hole collisions(*58*), or superconductivity(*59, 60*).

**Comparison with Shubnikov–de Haas (SdH) oscillations**

To further elucidate the band splitting in our system, we have measured SdH oscillations at higher magnetic field range. The results are summarized in Fig. 5. In all the density range including the electron side, we found beatings in the oscillations originating from the spin-orbit-coupled split bands (Fig. 5B). For quantitative analysis, we performed fast Fourier transforms (FFT) and extracted the frequency $f$ at which the spectra exhibit a peak (Fig. 5A) which is directly connected to the area of the corresponding Fermi surface by $f = nh/2e$ and $n = k_F^2/2\pi$ assuming broken spin degeneracy due to SOC. It can therefore be used to estimate the SOC strengths independently. Fig. 5C shows the result (we have selected the peaks that evolve continuously in density only) which shows a good agreement with the calculation using the SOC strengths along the circle $\lambda_{SOC} = 3.4 \pm 0.7$ meV (Fig. 5D). Interestingly, the fitting is slightly better (i.e., $\langle \delta B^2 \rangle$ is smaller) near $\lambda = 0$ which indicates the larger Rashba SOC in the system, consistent with our estimation from the analysis on the second focusing peak (Figs. 3C,D). The absolute value is however about 4 times smaller than those extracted from the TMF data (Figs. 3A,B), indicating that the TMF and SdH oscillations probe different electron dynamics.

To compare the results more directly, we have calculated $\Delta B_{SdH} = (2\hbar/|e|L)|k_{F1} - k_{F2}|$, the expected size of the splitting of the first TMF peak in $B$ using the $k_F$-values extracted from the SdH oscillations (Fig. 5C) and plotted the splitting measured in TMF, $\Delta B_{TMF}$ (extracted from Figs. 2A-C), together in Fig. 5E. As shown in the figure, $\Delta B_{TMF}$ remains larger than $\Delta B_{SdH}$ in all density range accessed in the experiment. Interestingly, we found that the similar behavior was observed in 2DEG systems with SOC(*19, 21*) where it was suggested(*21*) that there might be a SOC term, such as a linear-*k* term, that does not affect the total area of the Fermi surface by shifting a circle in one momentum direction. Since the SdH oscillation requires carriers to complete a full cyclotron motion while in TMF, they only make a half turn, it may be possible that one finds larger splitting in TMF than in SdH as seen experimentally. However, it is also possible that the SdH oscillations require relatively large magnetic fields to form Landau levels which can induce non-negligible Zeeman energy and may affect the spin-valley Zeeman and Rashba terms in Eq. (1) differently (see **Supplementary Materials** for more discussions). Although more studies are required to understand this discrepancy, our measurement shows that the behavior occurs not only in semiconductor heterostructures-based 2DEG systems(*19, 21*) but also in graphene when SOC exists. Therefore, it is likely that there is a fundamental origin behind this phenomenon.

**Comparison with previous studies**

In Fig. 5D, we also include all SOC strengths extracted from the previous measurements on monolayer graphene-TMDC heterostructures for comparison. Overall, the relaxation time analysis from weak anti-localization or spin-Hall effect measurements(*1, 32, 33, 37*) shows a large sample-to-sample variation (Fig. 5D). It can be from the fact that these measurements rely on the model to connect the spin relaxation process in the system with the SOC strength, which is sensitive to the sample-specific electron scattering process(*7, 37*). On the other hand, TMF and SdH oscillations directly probe the size of the Fermi surface, which can be compared with theoretically calculated band structures without considering details of the scattering processes. The recent study on SdH oscillation in monolayer graphene-WSe$_2$ heterostructures(*38*) has indeed shown a SOC strength $\lambda_{SOC} = 2.51$ meV close to ours (a dotted circle in Fig. 5D). Interestingly, the study on Landau level splitting(*61*), which is closely related to the SdH oscillations, also showed a similar SOC strength (a square in Fig. 5D). Moreover, in our TMF study, we found similar SOC strengths in two different samples (Figs. 3A,B). This further elaborates the benefits of carrying out the (ballistic) transport spectroscopy on understanding electronic properties of the system with SOC.

**Discussion**

In summary, we have successfully demonstrated the ballistic electron motion in graphene on WSe$_2$ by measuring TMF signals at different charge density, magnetic fields, and temperature. From the density and magnetic field dependence of the first focusing peaks (Fig. 2), we confirmed that there exists two split bands in the system as expected theoretically(*2*) and estimated the SOC strength of $\lambda_{SOC} = 13.0 \pm 4.7$ meV (Figs. 3A,B). More interestingly, by analysing the behavior of the second focusing peak that shows no splitting and by carrying out quantum transport simulations, we were able to learn that the Rashba SOC is likely dominant in our system (Figs. 3C,D). Both the presence of the band splitting and a stronger Rashba SOC are well reproduced in SdH oscillations measurements (Fig. 5) even though they showed a smaller SOC strength of $\lambda_{SOC} = 3.4 \pm 0.7$ meV. Similar discrepancy was found in other 2DEG with Rashba SOC(*19, 21*) which indicates that there is a fundamental reason behind it. This calls for new theory.

In addition to providing spectroscopic evidence of the spin-orbit-coupled bands, our work demonstrates that graphene on TMDCs can support ballistic transport that can be used not only to gain more insights of the microscopic electron process in the system but also to exploit various ballistic transport effects that are pronounced in graphene. It is particularly interesting as, in contrast to the existing studies on graphene spintronics(*7, 22, 62, 63*), TMF separates spin-up and spin-down carriers in real space. This enables detection and measurement of both spins independently, instead of only the majority one injected from magnetic contacts. This, therefore, offers an alternative venue for graphene spintronic applications(*7, 22, 62, 63*). The similar strategies can also be used to study other 2D materials or heterostructures with strong SOC, such as bilayer graphene-TMDC heterostructures(*10-13*), black phosphorus(*64*), and more, which will offer new understandings about the effect of SOC in these material systems.

**Materials and Methods**

**Sample fabrication**

The WSe$_2$, h-BN and graphene flakes were exfoliated from corresponding crystals onto silicon wafers and examined under an optical microscope. The flakes with suitable thicknesses and surfaces were selected and assembled onto highly doped silicon substrates with 285-nm-thick oxide, following the standard dry pick-up and transfer technique(*46*). After the assembly, the stacks were annealed at 250 °C for 2 hours in a tube furnace in Ar/H$_2$ forming gas. 1D electrical contacts were fabricated on the annealed sample through a standard electron-beam lithography and reactive-ion etching (CF$_4$/O$_2$ mixture gas with flow rates of 5/25 sccm, RF power: 60W), followed by electron beam evaporation of 5 nm Cr and 50 nm Au films. The devices were finally shaped into Hall bars by another electron-beam lithography and reactive-ion etching process.

**Electrical measurement**

Devices were measured in a 1.5 K cryogen-free variable temperature insert (VTI) with a superconducting magnet. The electrical signals were measured by applying a small low-frequency (17.777 Hz) AC current of 0.1–1 µA between the source and drain terminals and measuring the voltage drop between another two probes using a lock-in amplifier (Stanford Research SR830). The low-noise filters and amplifiers were used to detect small TMF signals. The back gate was controlled by Keithley 2400 source-meter.

**Acknowledgments**
**Funding:** The work is financially supported by the National Key R&D Program of China (2020YFA0309600) and by the University Grants Committee/Research Grant Council of Hong Kong SAR under schemes of Area of Excellence (AoE/P-701/20), ECS (27300819), and GRF (17300020, 17300521, 17309722). K.W. and T.T. acknowledge support from JSPS KAKENHI (Grant Numbers 19H05790, 20H00354, and 21H05233) and A3 Foresight by JSPS. N.W. acknowledges support from William Mong Institute of Nano Science and Technology. W.-H. K. and M.-H. L. gratefully acknowledge National Science and Technology Council of Taiwan (grant number: MOST 109-2112-M-006-020-MY3) for financial support and National Center for High-performance Computing (NCHC) for providing computational and storage resources.

**Author contributions:** DKK conceived and supervised the project. MHL supervised the theoretical part carried out by WHK. QR fabricated the samples and performed the measurements, assisted by HX. Some of the data were collected in NW's group with helps from ZY and XF. QR and WHK analyzed the data, and DKK and MHL interpreted them with input from all authors. TT and KW synthesized the hBN crystals. QR, HX, WHK, MHL, and DKK wrote the paper with input from all authors. All authors discussed the results.

**Competing interests:** The authors declare they have no competing interests.

**Figures and Tables**

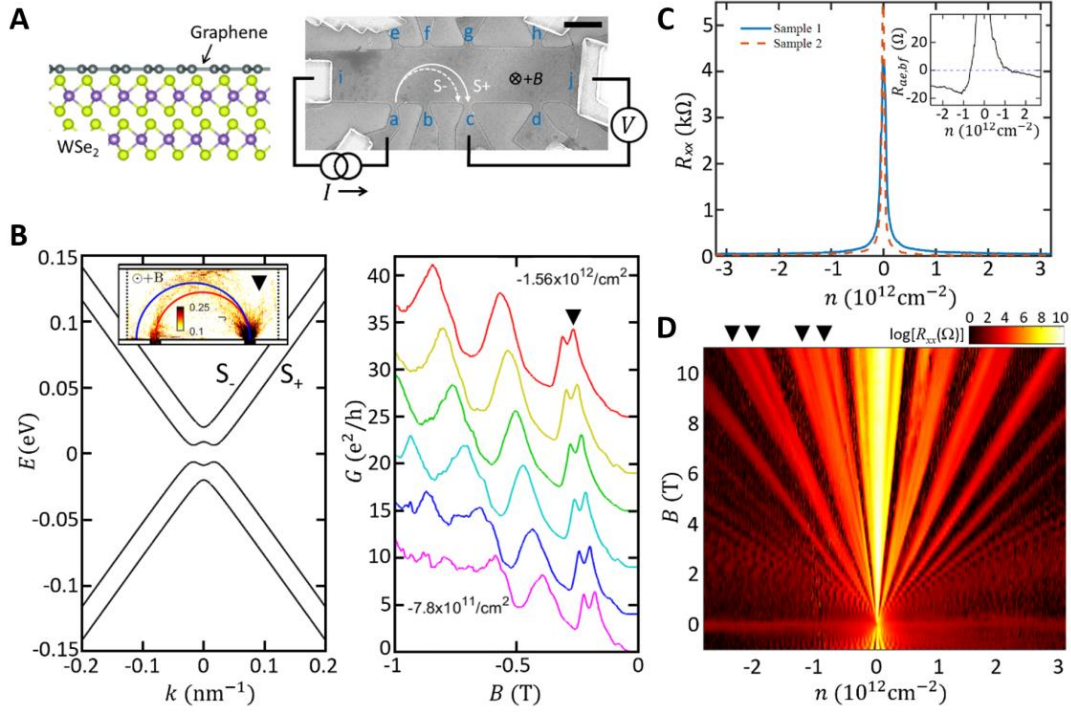

**Fig. 1. Sample characteristics and TMF measurement scheme.** (**A**) The schematic of monolayer graphene-multilayer $WSe_2$ heterostructures (left) and scanning electron microscope image of the device with a TMF measurement configuration (right; a scale bar: 2 µm). The two semicircles ($S_\pm$) on the right illustrate trajectories of the carriers at different spin-orbit-coupled bands $S_\pm$ shown in (B) under perpendicular magnetic field $B$. (**B**) The energy dispersion of graphene on $WSe_2$ derived from the effective Hamiltonian Eq. (1) in the main text (left) using SOC strengths of $\lambda = \lambda_R = 8.9$ meV ($\lambda_{SOC} \equiv \sqrt{\lambda^2 + \lambda_R{}^2} = 12.5$ meV), and the corresponding TMF spectra at 6 different hole densities, $n$ (in $10^{12}$ cm$^{-2}$) = -0.78, -0.93, -1.09, -1.24, -1.40, and -1.56 (from bottom to top) calculated from the tight-binding model (right). The inset on the left shows the simulated local current density map at the focusing peak marked by a down triangle on the right. (**C**) The carrier density ($n$) dependence of the four-probe resistances $R_{xx}$ of two different samples 1 (blue solid line) and 2 (red broken line) measured at 1.5 K. Both exhibits a sharp resistance peak at zero density, indicating high device quality. The inset shows the non-local Hall resistance $R_{ae,bf}(\equiv V_{bf}/I_{ae})$, exhibiting a large negative signal on the hole side originating from the ballistic transport. (**D**) The Landau fan—the $\log(R_{xx})$ as a function of $n$ and $B$—plotted in a color scale (the darker color corresponds to the lower resistance), showing high-quality quantum Hall effect measured at 1.5 K. On the hole side, the broken-symmetry states begin to appear at ~3 T (indicated by black down-triangles) which indicates higher quality on the hole side, consistent with the large negative $R_{ae,bf}$ on the hole side shown in the inset of (C).

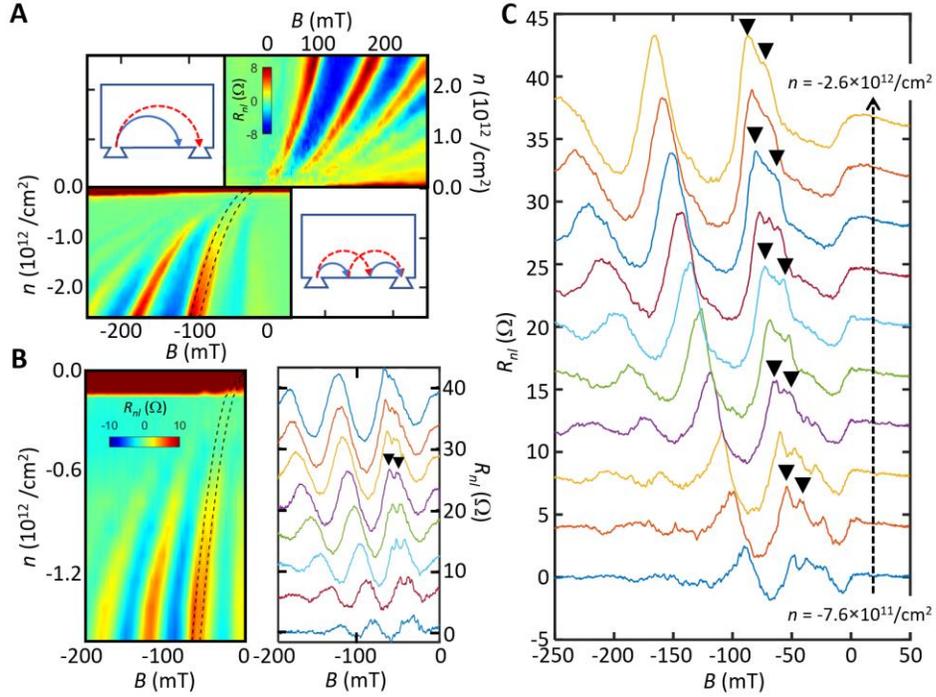

**Fig. 2. TMF spectra.** (**A**) Color-scale maps of TMF signal $R_{nl}(B, n)$ measured in sample 1 at 1.5 K (top right: electron side; bottom left: hole side). The broken lines show the theoretically calculated focusing peaks at $\lambda_{SOC} = 13.9$ meV. Inset: carrier trajectories for the first and second focusing peaks (top left and bottom right, respectively). (**B**) TMF map and the corresponding 1D cuts measured in sample 2 at 1.5 K. The broken line shows the theoretically calculate focusing peaks at $\lambda_{SOC} = 12.0$ meV. (**C**) 1D cuts of the data from the sample 1 shown in (A). The black down-triangles in (B) and (C) mark some of the two split peaks for guidance.

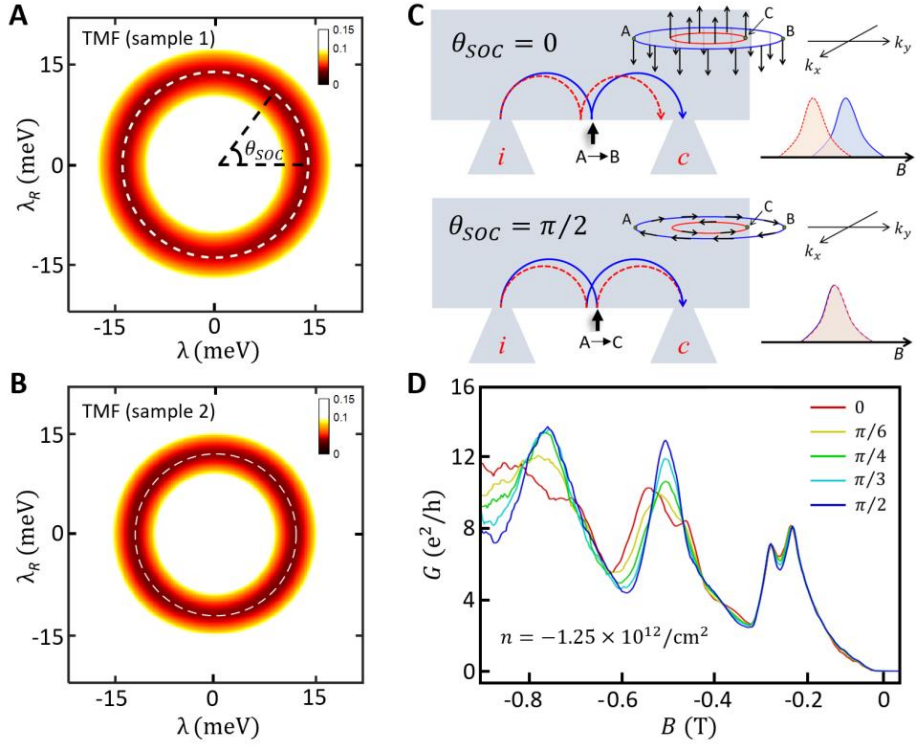

**Fig. 3. Analysis on TMF signal. (A-B)** The color-scale map of the average difference $\langle \delta B^2 \rangle \equiv \sum[(\Delta B_+/B_0)^2 + (\Delta B_-/B_0)^2]/N$ as a function of $\lambda$ and $\lambda_R$ from the sample 1 and 2, respectively (*N*: number of data used). $\Delta B_\pm$ is the difference between the predicted focusing peak positions from the simulation for certain $(\lambda, \lambda_R)$ and the real peak positions measured in the experiment for the band $S_\pm$, whereas $B_0$ is the half of the maximum splitting observed. Thus, the smaller $\langle \delta B^2 \rangle$ (darker in the map) indicates a better agreement. The best-fit value is drawn by a dashed white circle. We use a criterion $\langle \delta B^2 \rangle \leq 0.1$ to extract the SOC strengths of $\lambda_{SOC} = 13.9 \pm 4.0$ meV and $12.0 \pm 3.5$ meV, for sample 1 and 2 respectively. $\theta_{SOC}$ in (A) is defined as $\cos^{-1}(\lambda/\lambda_{SOC})$. **(C)** Comparison of the electron trajectories for the second focusing peak for the two cases when there is only the spin-valley Zeeman term (top, $\theta_{SOC} = 0$) and when only Rashba term exists (bottom, $\theta_{SOC} = \pi/2$). The schematic band structures with spin configurations and the shapes of the resulting focusing peaks for each case are shown in the inset. Due to the spin conservation, the electron at the edge (the state "A") will be backscattered to "B" when $\theta_{SOC} = 0$, leading to the splitting in the second peak, whereas when $\theta_{SOC} = \pi/2$, it will be transferred to "C" in a different band. **(D)** The calculated TMF spectra with varying $\theta_{SOC}$ at $n = -1.25 \times 10^{12}$ cm$^{-2}$ when the overall SOC strength $\lambda_{SOC} = 10$ meV. One can clearly see that the positions of the first focusing peaks remains the same while the second peak shows multiple peaks near $\theta_{SOC} = 0$.

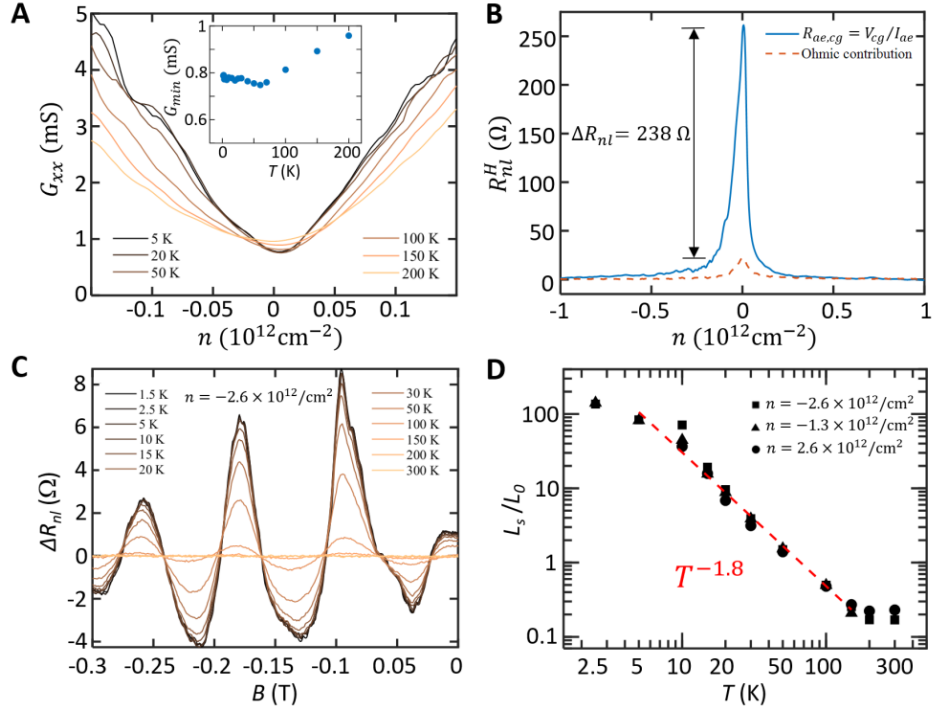

**Fig. 4. Temperature dependence measurements.** (**A**) Temperature dependence of the local four-terminal conductance $G_{xx} = 1/R_{xx}$ as a function of charge density, exhibiting a weak temperature dependence of the minimum conductance $G_{min}$ at zero density as magnified in the inset. (**B**) Non-local Hall resistance $R_{ae,cg}$ as a function of carrier density $n$ (blue solid line) compared with the calculated Ohmic contribution (red broken line), consistent with the spin Hall effect. (**C**) Temperature dependence of the TMF spectra at $n = -2.6 \times 10^{12}$ cm$^{-2}$. The smooth backgrounds are extracted by a Gaussian filter with a full width at half maximum of 0.2 T, which is larger than the oscillation period of TMF signals. (**D**) The relative scattering lengths (calculated from areas below the first focusing peaks) at $n = -2.6 \times 10^{12}$ cm$^{-2}$ (square), $n = -1.3 \times 10^{12}$ cm$^{-2}$ (triangle), and $n = 2.6 \times 10^{12}$ cm$^{-2}$ (circle) as a function of temperature plotted in a log scale, which follows the $T^{-1.8}$ dependence, indicated by the dashed red line. See the main text for more discussions.

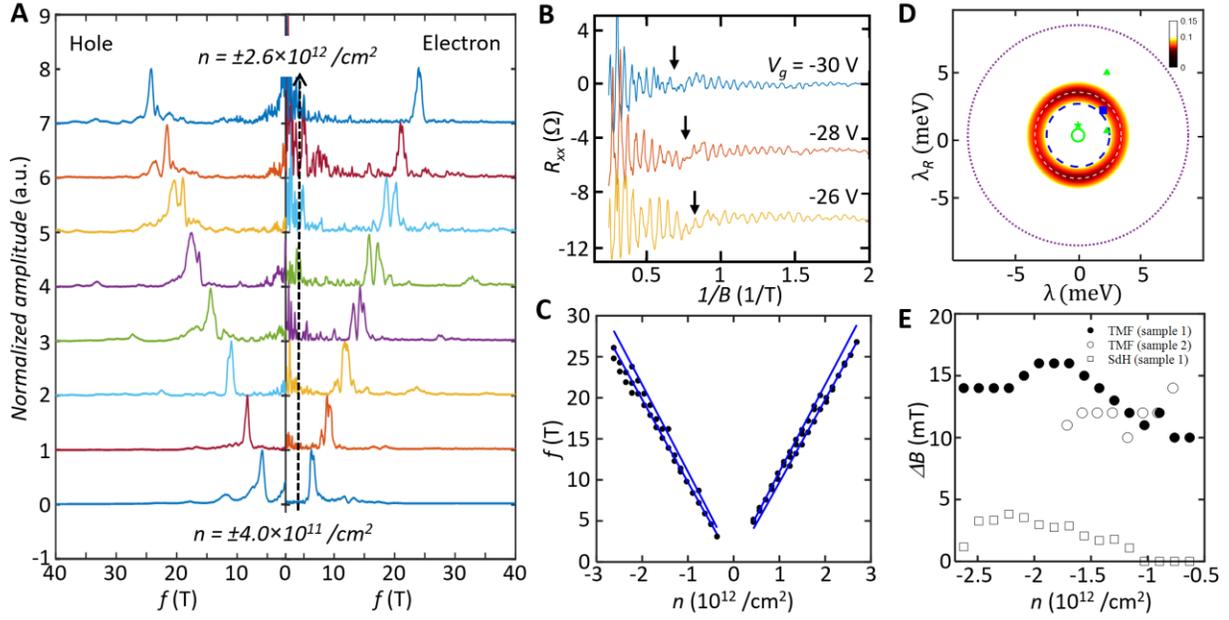

**Fig. 5. SdH oscillations.** (**A**) The FFT spectra derived from the SdH oscillation curves as a function of carrier density $n$, for both electron and hole side. (**B**) Representative SdH oscillation curves, which clearly show the beating patterns (marked by arrows; curves are vertically offset for clarity). (**C**) The frequency peak positions extracted from the FFT spectra at different carrier densities (see **Supplementary Materials**). The solid lines show the fitting with calculated band structures using $\lambda_{SOC} = 3.4$ meV. (**D**) The color-scale map of the average difference $\langle \delta B^2 \rangle$ in $\lambda$ and $\lambda_R$ from the SdH oscillations with the best fit value drawn by dashed white circle (see the caption of Figs. 3A,B for details). Using the same criterion used in Figs. 3A,B $\langle \delta B^2 \rangle \leq 0.1$, we get $\lambda_{SOC} = 3.4 \pm 0.7$ meV. Note that around the circle near $\lambda = 0$, the color becomes darker which indicates stronger Rashba SOC in the system consistent with discussions in Figs. 3C,D. The values from the previous studies are included for comparison: a dotted purple circle from spin Hall effect(*1*), a dashed blue circle from SdH oscillations(*38*), a square from Landau level splitting(*61*), and circle(*32*), star(*33*), and two filled green triangles(*37*) (two different values are obtained from different spin relaxation mechanisms) from weak anti-localization measurements. Some studies(*1, 32, 33, 61*) used $\lambda_R/2$ and/or $\lambda/2$ in the Hamiltonian Eq. (1), so we divided the values by half in the plot. (**E**) The splitting of the first TMF peak in $B$ ($\Delta B_{TMF}$) at different charge densities extracted from Fig. 2 compared with the predicted $\Delta B$ calculated from the SdH oscillations ($\Delta B_{SdH}$). Over the whole density range, the $\Delta B_{SdH}$ remains smaller than the $\Delta B_{TMF}$.

# Supplementary Materials

**Section 1: Quantum transport simulation**

Current density maps (the inset of Fig. 1B and Fig. S1E) and conductance spectra (Fig. 1B, right) were simulated using Kwant(*44*), an open-source python package for quantum transport simulations based on tight-binding models. To model transport in graphene on TMDCs, we adopt the effective tight-binding Hamiltonian(*6*)

$$H = H_0 + H_1 + H_R + H_{vZ} + H_{PIA}, \quad (S1)$$

where the first two terms are spin-independent and the rest describe spin-orbit couplings of different origins. The first term, $H_0 = \sum_{\langle i,j \rangle, \sigma} t c_{i\sigma}^\dagger c_{j\sigma}$, is composed of nearest-neighbor kinetic hopping of strength $t$, and is typically used to describe a spinless graphene lattice. Here, $i, j$ are lattice site indices, $\sigma$ is the spin index, $c_{i\sigma}^\dagger$ ($c_{j\sigma}$) is the creation (annihilation) operator that creates (annihilates) an electron of spin $\sigma$ at site $i$, and $\sum_{\langle i,j \rangle}$ sums over site indices that are nearest to each other. Except $H_0$, all the rest of the terms in Eq. (S1) arise from the effect of the neighboring TMDC lattice. First, the symmetry between the sublattices A and B of the graphene lattice is broken, giving rise to an effective energy difference experienced by electrons on atoms of the two sublattices, which can be described by

$$H_1 = \sum_{i,\sigma} \xi_{o_i} \Delta c_{i\sigma}^\dagger c_{i\sigma},$$

where $o_i$ is the sublattice index of site $i$, $\xi_{o_i} = +1 \,(-1)$ when $o_i = A \,(o_i = B)$, and $\Delta$ characterizes the strength of such a staggered potential energy. The rest of the three terms in Eq. (2) include the Rashba spin-orbit coupling,

$$H_R = \frac{2i}{3} \sum_{\langle i,j \rangle} \sum_{\sigma, \sigma'} c_{i\sigma}^\dagger c_{j\sigma'} \left[ \lambda_R (\hat{s} \times \hat{d}_{ij})_z \right]_{\sigma, \sigma'},$$

where $\lambda_R$ the Rashba coupling strength, $\hat{s} = (\hat{s}_x, \hat{s}_y, \hat{s}_z)$ is a vector of Pauli matrices acting on spin, and $\hat{d}_{ij}$ is a unit vector pointing from site j to $i$, the valley-Zeeman term,

$$H_{vZ} = \frac{i}{3\sqrt{3}} \sum_{\langle\langle i,j \rangle\rangle} \sum_{\sigma, \sigma'} c_{i\sigma}^\dagger c_{j\sigma} \left[ \lambda_I^{o_i} v_{ij} \hat{s}_z \right]_{\sigma, \sigma'},$$

where $\sum_{\langle\langle i,j \rangle\rangle}$ sums over site indices $i, j$ that are second nearest to each other, the sign factor $v_{ij} = +1 \,(-1)$ when the resulting hopping path is counterclockwise (clockwise), and $\lambda_I^{o_i}$ is the sublattice-resolved valley-Zeeman coupling strength, and finally the pseudospin-inversion-asymmetry term $H_{PIA}$ that does not influence the band structure at $K$ and $K'$. Neglecting $H_{PIA}$ and setting $\lambda_I^A = -\lambda_I^B = \lambda$ for both sublattices $o_i = A, B$, the eigenenergy of tight-binding model Hamiltonian Eq. (3) is given by(5)

$$E_{\mu,\nu}(k) = \mu\sqrt{(\Delta^2 + \lambda^2 + 2\lambda_R^2 + \hbar^2 v_F^2 k^2) + 2\nu\sqrt{(\lambda_R^2 - \lambda\Delta)^2 + (\lambda^2 + \lambda_R^2)\hbar^2 v_F^2 k^2}}$$

with $\mu, \nu = \pm 1$, whose low-$k$ expansion simplifies to the eigenenergy of Eq. (1) of the main text, which is adopted in several previous studies(2,4,5,38). Note that to account for micron-sized graphene systems, we have adopted the scaled tight-binding model(45) which is compatible with the spin-orbit coupling terms as remarked in the recent study(65). All quantum transport simulations presented here are based on the scaling factor $s_F = 8$.

### Section 2: Semi-classical ray tracing

In order to check if the absence of the splitting in the second focusing peak (Fig. 1B right and Figs. 2A-C) is due to the inter-band scattering at the edge, we have employed the semi-classical ray tracing by solving the following equation of motion(66):

$$\begin{cases} \dfrac{d\vec{v}}{dt} = \dfrac{1}{\hbar}\vec{\nabla}_{\vec{k}}E(\vec{k}) \\ \dfrac{d\vec{k}}{dt} = -\dfrac{e}{\hbar}\vec{v} \times \vec{B} \end{cases},$$

where one can manually add or remove scattering conditions like the inter-band transition at the edge. Fig. S1 shows the results that clearly prove that the absence of the splitting in the second focusing peak is from the scattering between the two bands $S_+$ and $S_-$ at the sample edge.

### Section 3: Discussions on TMF and SdH oscillations

As discussed briefly in the main text, in TMF, the electrons make only half of the cyclotron motion while in SdH oscillations, it needs to make a full circle without losing its phase coherence. This can give rise to several differences in the two phenomena. First, the TMF is more sensitive to the electron scattering than the SdH oscillations as TMF is from the ballistic motion of electrons that can be destroyed by elastic scattering. Thus, in our sample, we could see the splitting in TMF focusing peak only on the hole side where we find higher sample quality while in SdH oscillations we found no obvious differences in both hole and electron side. Second, as electrons need to make a full cyclotron motion, SdH oscillations probe the total Fermi surface area only, whereas the TMF probes the partial trajectory along the Fermi surface. This can make a difference when the band structure is shifted in one momentum direction while keeping its area. This might be the reason why the studies on 2DEG with SOC including ours showed different splitting in TMF and SdH oscillations. Lastly, the SdH oscillations generally occur at larger magnetic fields than the TMF. This may result in a stronger effect of the Zeeman energy that couples to out-of-plane spin components. It could thus affect the spin-valley Zeeman and Rashba terms in Eq. (1) differently, that may lead to different band splitting in SdH oscillations compared with TMF. Nonetheless, the microscopic process that governs TMF and SdH oscillations is different and our study, together with other studies on 2DEG with SOC(*19, 21*), shows that it might be important to consider their differences when analysing the relevant experimental results.

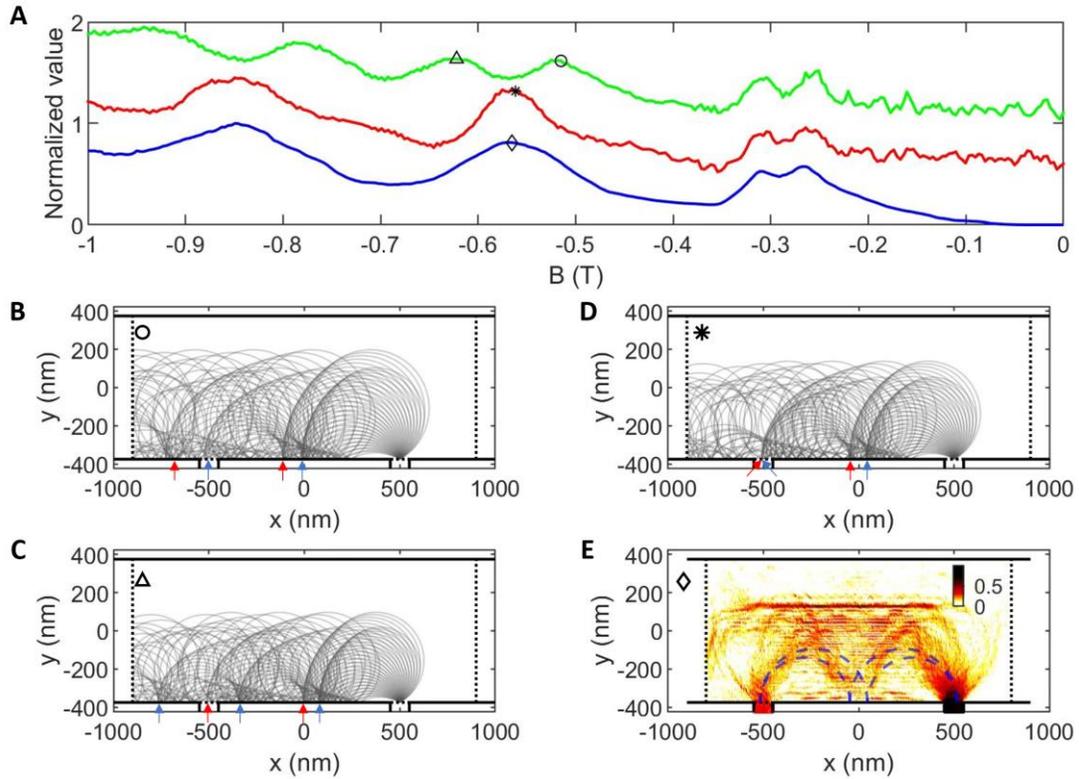

**Fig. S1. Inter-band transition simulation for the second focusing peak.** (**A**) Top green and middle red curves: normalized two-point TMF conductance calculated by semi-classical ray tracing without or with inter-band transition, respectively. Bottom blue curve: normalized conductance from quantum transport simulation. As expected, the second focusing peak shows splitting when the inter-band transition is switched off (top green curve). Moreover, the positions of the focusing peaks calculated with inter-band transition (middle red curve) match well with those from Kwant simulation (bottom blue curve), indicating that the observed absence of the splitting in the second focusing peak (Figs. 2A-C) is from the inter-band transition. (**B-D**), The calculated carrier trajectories at the focusing peaks marked by circle, triangle, and star in (A), respectively. Red and blue arrows indicate the positions of the scattering at the edge for carriers at $S_{\pm}$ bands, respectively. (**E**) Local current density distribution at the focusing peak marked by diamond from quantum transport simulation. Dashed lines illustrate the carrier trajectories.